\documentclass[10pt,conference]{IEEEtran}
\usepackage{psfrag}
\usepackage{subfigure}
\usepackage{cite}

\usepackage{array,multirow}
\usepackage{graphicx,color}  
\usepackage{psfrag}    
\usepackage{subfigure} 
\usepackage{amsmath}   
\usepackage{amssymb}   
\usepackage{eucal}     

\usepackage{pxfonts}
\usepackage{dsfont}

\DeclareMathAlphabet{\eurm}{U}{eur}{m}{n}
\DeclareMathAlphabet{\mathbsf}{OT1}{cmss}{bx}{n}
\DeclareMathAlphabet{\mathssf}{OT1}{cmss}{m}{sl}
\DeclareMathAlphabet{\mathcsf}{OT1}{cmss}{sbc}{n}

\DeclareSymbolFont{bsfletters}{OT1}{cmss}{bx}{n}  
\DeclareSymbolFont{ssfletters}{OT1}{cmss}{m}{n}
\DeclareMathSymbol{\bsfGamma}{0}{bsfletters}{'000}
\DeclareMathSymbol{\ssfGamma}{0}{ssfletters}{'000}
\DeclareMathSymbol{\bsfDelta}{0}{bsfletters}{'001}
\DeclareMathSymbol{\ssfDelta}{0}{ssfletters}{'001}
\DeclareMathSymbol{\bsfTheta}{0}{bsfletters}{'002}
\DeclareMathSymbol{\ssfTheta}{0}{ssfletters}{'002}
\DeclareMathSymbol{\bsfLambda}{0}{bsfletters}{'003}
\DeclareMathSymbol{\ssfLambda}{0}{ssfletters}{'003}
\DeclareMathSymbol{\bsfXi}{0}{bsfletters}{'004}
\DeclareMathSymbol{\ssfXi}{0}{ssfletters}{'004}
\DeclareMathSymbol{\bsfPi}{0}{bsfletters}{'005}
\DeclareMathSymbol{\ssfPi}{0}{ssfletters}{'005}
\DeclareMathSymbol{\bsfSigma}{0}{bsfletters}{'006}
\DeclareMathSymbol{\ssfSigma}{0}{ssfletters}{'006}
\DeclareMathSymbol{\bsfUpsilon}{0}{bsfletters}{'007}
\DeclareMathSymbol{\ssfUpsilon}{0}{ssfletters}{'007}
\DeclareMathSymbol{\bsfPhi}{0}{bsfletters}{'010}
\DeclareMathSymbol{\ssfPhi}{0}{ssfletters}{'010}
\DeclareMathSymbol{\bsfPsi}{0}{bsfletters}{'011}
\DeclareMathSymbol{\ssfPsi}{0}{ssfletters}{'011}
\DeclareMathSymbol{\bsfOmega}{0}{bsfletters}{'012}
\DeclareMathSymbol{\ssfOmega}{0}{ssfletters}{'012}

\newtheorem{theorem}{\textbf{Theorem}}

\newtheorem{proposition}{\textbf{Proposition}}
\newtheorem{definition}{\textbf{Definition}}

\newtheorem{remark}{Remark}
\newcommand{\dv}{\mathbf} 
\newcommand{\mc}{\mathcal} 
\newcommand{\qed}{\hfill \ensuremath{\Box}}

\begin{document}
\fontsize{9.4}{11.25pt}\selectfont

\title{Wyner-Ziv Type Versus Noisy Network Coding For a State-Dependent MAC\\}

\author{Abdellatif Zaidi$\:^{\dagger}$ \qquad Pablo Piantanida$\:^{\nmid}$ \qquad Shlomo Shamai (Shitz)$\:^{\ddagger}$\vspace{0.3cm}\\
$^{\dagger}$ Universit\'e Paris-Est Marne La Vall\'ee, Champs-sur-Marne 77454, France\\
$^{\nmid}$ Department of Telecommunications, SUPELEC, 91192 Gif-sur-Yvette, France\\
$^{\ddagger}$ Department of EE, Technion-Israel Institute of Technology, Haifa, Israel\\
abdellatif.zaidi@univ-mlv.fr, pablo.piantanida@supelec.fr, sshlomo@ee.technion.ac.il
}

\maketitle

\begin{abstract}
We consider a two-user state-dependent multiaccess channel in which the states of the channel are known non-causally to one of the encoders and only strictly causally to the other encoder. Both encoders transmit a common message and, in addition, the encoder that knows the states non-causally transmits an individual message. We find explicit characterizations of the capacity region of this communication model in both discrete memoryless and memoryless Gaussian cases. 
The analysis also reveals optimal ways of exploiting the knowledge of the state only strictly causally at the encoder that sends only the common message when such a knowledge is beneficial. The encoders collaborate to convey to the decoder a lossy version of the state, in addition to transmitting the information messages through a generalized Gel'fand-Pinsker binning. Particularly important in this problem are the questions of 1) optimal ways of performing the state compression and 2) whether or not the compression indices should be decoded uniquely. We show that both compression \`a-la noisy network coding, i.e., with no binning, and compression using Wyner-Ziv binning are optimal. The scheme that uses Wyner-Ziv binning shares elements with Cover and El Gamal original compress-and-forward, but differs from it mainly in that backward decoding is employed instead of forward decoding and the compression indices are not decoded uniquely. Finally, by exploring the properties of our outer bound, we show that, although not required in general, the compression indices can in fact be decoded uniquely essentially without altering the capacity region, but at the expense of larger alphabets sizes for the auxiliary random variables.
\end{abstract}

\section{Introduction}\label{secI}

Advances in the study of the effect of strictly causal states in multiuser channels are rather very recent and concern mainly multiple access scenarios. In \cite{LS10a}, Lapidoth and Steinberg study a two-encoder multiple access channel with independent messages and states known causally at the encoders. They show that the strictly causal state sequence can be beneficial, in the sense that it increases the capacity for this model. This result is reminiscent of Dueck's proof \cite{D80} that feedback can increase the capacity region of some broadcast channels. In accordance with \cite{D80}, the main idea of the achievability result in \cite{LS10a} is a block Markov coding scheme in which the two users collaborate to describe the state to the decoder by sending cooperatively a compressed version of it. As noticed in \cite{LS10a}, although some non-zero rate that otherwise could be used to transmit pure information is spent in describing the state to the decoder, the net effect can be an increase in the capacity. In \cite{LS10b}, they show that strictly causal state information is beneficial even if the channel is controlled by two independent states each known to one encoder strictly causally. In this case, each encoder can help the other encoder transmit at a higher rate by sending a compressed version of its state to the decoder. In \cite{LSY10}, Li, Simeone and Yener improve the results of \cite{LS10a,LS10b} and extend them to the case of multiple encoders.  The achievability results in \cite{LSY10} are inspired by the noisy network coding scheme of \cite{H-LKGC11} and, unlike \cite{LS10a,LS10b}, do not use Wyner-Ziv binning \cite{WZ76} for the compression of the state. In a very recent contribution \cite{LS11a}, Lapidoth and Steinberg derive a new inner bound on the capacity region for the case of a single state governing the multiaccess channel. They also prove that the inner bound of \cite{LSY10} for the case of two independent states each known strictly causally to one encoder can indeed be strictly better than previous bounds in \cite{LS10a,LS10b} -- a result which is conjectured previously by Li, Simeone and Yener in \cite{LSY10}.
\vspace{-0.2cm}

\begin{figure}[htpb]
\centering
\includegraphics[width=0.9\linewidth]{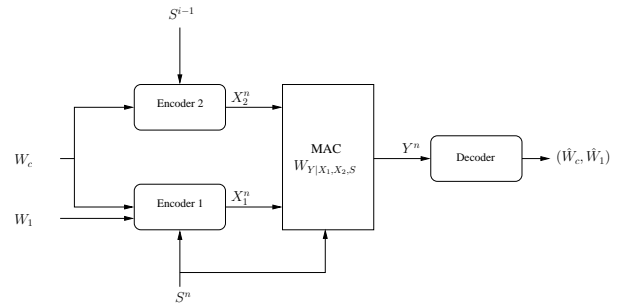}
\vspace{-0.3cm}
\caption{State-dependent MAC with degraded message sets and states known noncausally at the encoder that sends both messages and only strictly causally at the other encoder.}
\label{ModelForMACwithAsymmetricCSI}
\end{figure}

\vspace{-0.2cm}

In this paper, which generalizes our former work \cite{ZPS11a}, we study a two-user state-dependent multiple access channel with the channel states known non-causally at one encoder and only strictly causally at the other encoder. The decoder is not aware of the channel states. As shown in Figure~\ref{ModelForMACwithAsymmetricCSI}, both encoders transmit a common message and, in addition, the encoder that knows the states non-causally transmits an individual message. More precisely, let $W_c$ and $W_1$ denote the common message and the individual message to be transmitted in, say, $n$ uses of the channel; and $S^n=(S_1,\hdots,S_n)$ denote the state sequence affecting the channel during this time. At time $i$, Encoder 1 knows the complete sequence $S^n=(S_1,\hdots,S_{i-1},S_i,\hdots,S_n)$ and sends $X_{1i}=\phi_1(W_c,W_1,S^n)$, and Encoder 2 knows \textit{only} $S^{i-1}=(S_1,\hdots,S_{i-1})$ and sends $X_{2i}=\phi_{2,i}(W_c,S^{i-1})$ -- the functions $\phi_1$ and $\phi_{2,i}$ are some encoding functions. In this paper, we establish the capacity region of this state-dependent MAC model. As our analysis will show, this requires, among others, understanding the role of the strictly causal part of the state that is revealed to Encoder 2. We show that both compression of the state \`a-la noisy network coding, i.e., with no binning, and compression using Wyner-Ziv binning with or without non-unique decoding of the compresion indices are optimal. The question of whether non-unique decoding is beneficial in Wyner-Ziv type networks is important. Related very recent works can be found in \cite{KH11a,WX11a}.

\vspace{-0.2cm}

\section{System Model and Definitions}\label{secII}

We consider a stationary memoryless state-dependent MAC $W_{Y|X_1,X_2,S}$  whose output $Y \in \mc Y$ is controlled by the channel inputs $X_1 \in \mc X_1$ and $X_2 \in \mc X_2$ from the encoders and a channel state $S \in \mc S$ drawn according to a memoryless probability law $Q_S$. We assume that the channel state $S^n$ is known non-causally at Encoder 1, i.e., beforehand, at the beginning of the transmission block. Encoder 2 knows the channel states only strictly-causally; that is, at time $i$, it knows the states only up to time $i-1$, $S^{i-1}=(S_1,\hdots,S_{i-1})$.

 Both encoders transmit a common message $W_c$ and, in addition, Encoder 1 transmits also an individual message $W_1$. We assume that $W_c$ and $W_1$ are independent random variables drawn uniformly from the sets $\mc W_c=\{1,\cdots,M_c\}$ and  $\mc W_1=\{1,\cdots,M_1\}$, respectively. The sequences $X_{1}^n$ and $X_{2}^n$ from the encoders are sent across a state-dependent multiaccess channel modeled as a memoryless conditional probability distribution $W_{Y|X_1,X_2,S}$. The joint probability mass function on ${\mc W_c}{\times}{\mc W_1}{\times}{\mc S^n}{\times}{\mc X^n_1}{\times}{\mc X^n_2}{\times}{\mc Y^n}$ is given by
\vspace{-0.3cm}
\begin{align}
P(w_c,w_1,s^n,x^n_1,x^n_2,y^n) &= P(w_c)P(w_1)\prod_{i=1}^{n} \big[Q_S(s_i)P(x_{1,i}|w_c,w_1,s^n)\nonumber\\
&\hspace{-0.5cm}{\cdot}P(x_{2,i}|w_c,s^{i-1})W_{Y|X_1,X_2,S}(y_i|x_{1,i},x_{2,i},s_i)\big].
\end{align}
The receiver guesses the pair $(\hat{W}_c,\hat{W}_1)$ from the output $Y^n$.

\begin{definition}
For positive integers $n$, $M_c$ and $M_1$, an $(M_c,M_1,n,\epsilon)$ code for the multiple access channel with states known noncausally at one encoder and only strictly causally at the other encoder consists of a mapping
\begin{align}
\phi_1: \mc W_c{\times}\mc W_1{\times}\mc S^n \longrightarrow \mc X^n_1
\label{EncodingFunction__Encoder1}
\end{align}
at Encoder 1, a sequence of mappings
\begin{align}
\phi_{2,i}: \mc W_c{\times}\mc S^{i-1} \longrightarrow \mc X_2, \quad i=1,\hdots,n
\label{EncodingFunction__Encoder2}
\end{align}
at Encoder 2, and a decoder map
\begin{align}
\psi : \mc Y^n \longrightarrow \mc W_c{\times}\mc W_1
\label{DecodingFunction}
\end{align}
such that the average probability of error is bounded by $\epsilon$,
\begin{equation}
P_e^n = \mathbb{E}_{S}\big[\mathrm{Pr}\big(\psi(Y^n)\neq (W_c,W_1)|S^n=s^n\big)\big] \leq \epsilon.
\end{equation}
The rate of the common message and the rate of the individual message are defined as
\begin{align}
&R_c = \frac{1}{n}\log M_c \qquad \text{and} \qquad R_1 = \frac{1}{n}\log M_1,
\end{align}
respectively.
\end{definition}

A rate pair $(R_c,R_1)$ is said to be achievable if for every $\epsilon > 0$ there exists an $(2^{nR_c},2^{nR_1},n,\epsilon)$ code for the channel $W_{Y|X_1,X_2,S}$.  The capacity region of the considered state-dependent MAC is defined as the closure of the set of achievable rate pairs.

Due to space limitation, the results of this paper are either outlined only or mentioned without proofs. Detailed proofs, converse proofs, the characterization of the capacity region in the Gaussian case as well as other results and discussions can be found in \cite{ZPS12b}.

\vspace{-0.1cm}

\section{Discrete Memoryless Case}\label{secIII}

In this section, it is assumed that $\mc S, \mc X_1, \mc X_2$ are finite.

\vspace{-0.1cm}

\subsection{Capacity Region}\label{secIII_subsecA}

Let $\mc P$ stand for the collection of all random variables $(S,U,V,X_1,X_2,Y)$ such that $U$, $V$, $X_1$ and $X_2$ take values in finite alphabets $\mc U$, $\mc V$, $\mc X_1$ and $\mc X_2$, respectively, and
\begin{subequations}
\begin{align}
&P_{S,U,V,X_1,X_2,Y}(s,u,v,x_1,x_2,y) \nonumber\\
&\hspace{0.6cm}= P_{S,U,V,X_1X_2}(s,u,v,x_1,x_2)W_{Y|X_1,X_2,S}(y|x_1,x_2,s)\\
&P_{S,U,V,X_1,X_2}(s,u,v,x_1,x_2) \nonumber\\
&\hspace{0.6cm}= Q_S(s)P_{X_2}(x_2)P_{V|S,X_2}(v|s,x_2)P_{U,X_1|S,V,X_2}(u,x_1|s,v,x_2)\\
&\sum_{u,v,x_1,x_2}P_{S,U,V,X_1,X_2}(s,u,v,x_1,x_2) \:= \: Q_S(s).
\end{align}
\label{MeasureForCapacityRegionDiscreteMemorylessChannel}
\end{subequations}

The relations in \eqref{MeasureForCapacityRegionDiscreteMemorylessChannel} imply that $(U,V) \leftrightarrow (S,X_1,X_2) \leftrightarrow Y$ is a Markov chain, and $X_2$ is independent of $S$.

Define $\mc C$ to be the set of all rate pairs $(R_c,R_1)$ such that
\begin{align}
R_1 \: &\leq \: I(U;Y|V,X_2)-I(U;S|V,X_2) \nonumber\\
R_c+ R_1 \: &\leq \: I(U,V,X_2;Y)-I(U,V,X_2;S)\nonumber\\
&\hspace{2cm} \text{for some}\:\: (S,U,V,X_1,X_2,Y) \in \mc P.
\label{CapacityRegionDiscreteMemorylessChannel}
\end{align}

The following proposition states some properties of $\mc C$.

\begin{proposition}\label{Proposition__Properties__of__CapacityRegion}

{\color{white} (properties of capacity region)}

\begin{itemize}
\item[1.] The set $\mc C$ is convex.
\item[2.] To exhaust $\mc C$, it is enough to restrict $\mc V$ and $\mc U$ to satisfy
\begin{subequations}
\begin{align}
\label{BoundsOnCardinalityOfAuxiliaryRandonVariableV__CapacityRegion__DiscreteMemorylessChannel}
&|\mc V| \leq |\mc S||\mc X_1||\mc X_2|+1\\
&|\mc U| \leq \Big(|\mc S||\mc X_1||\mc X_2|+1\Big)|\mc S||\mc X_1||\mc X_2|.
\label{BoundsOnCardinalityOfAuxiliaryRandonVariableU__CapacityRegion__DiscreteMemorylessChannel}
\end{align}
\label{BoundsOnCardinalityOfAuxiliaryRandonVariables__CapacityRegion__DiscreteMemorylessChannel}
\end{subequations}
\end{itemize}
\end{proposition}

\vspace{-0.4cm}

\noindent As stated in the following theorem, the set $\mc C$ characterizes the capacity region of the state-dependent discrete memoryless MAC model that we study.

\vspace{0.2cm}

\begin{theorem}\label{Theorem__CapacityRegionDiscreteMemorylessChannel}
The capacity region of the multiple access channel with states known only strictly causally at the encoder that sends the common message and non-causally at the encoder that sends both messages is given by $\mc C$.
\end{theorem}

\vspace{0.3cm}

\begin{remark}\label{remark1}
The proof of achievability of Theorem~\ref{Theorem__CapacityRegionDiscreteMemorylessChannel} is based on a block-Markov coding scheme in which a lossy version of the state is conveyed to the decoder, in the spirit of \cite{LS10a,LS10b,LS11a}, in addition to a generalized Gel'fand-Pinsker binning for the transmission of the information messages \cite{GP80}. However, unlike \cite{LS10a,LS10b} and \cite{LS11a} where Wyner-Ziv compression is utilized for the transmission of the lossy version of the state, here, inspired by the noisy network coding scheme of \cite{H-LKGC11}, at each block the compression index of the state of the previous block is sent using standard rate distortion, not Wyner-Ziv binning \cite{WZ76}. Also, unlike \cite{LS10a,LS10b} and \cite{LS11a} where every information message is divided into blocks and different submessages are sent over these blocks and then decoded one at a time using the same codebook as in the original compress-and-forward scheme by Cover and El Gamal \cite{CG79}, here the \textit{entire} common message and the \textit{entire} individual message are transmitted over \textit{all} blocks using codebooks that are generated independently, one for each block, and the decoding is performed simultaneously using all blocks as in \cite{H-LKGC11}. At the end of the transmission, the receiver uses the outputs of all blocks to perform simultaneous decoding of the information common and individual messages, without uniquely decoding the compression indices. \qed
\end{remark}

\textbf{Proof of Achievability:}

The transmission takes place in $B$ blocks. The common message $W_c$ and the individual message $W_1$ are sent over \textit{all} blocks. We thus have $B_{W_c}=nB{R_c}$, $B_{W_1}=nB{R_1}$, $N=nB$, $R_{W_c}=B_{W_c}/N=R_c$ and $R_{W_1}=B_{W_1}/N=R_1$, where $B_{W_c}$ is the number of common message bits, $B_{W_1}$ is the number of individual message bits, $N$ is the number of channel uses and $R_{W_c}$ and $R_{W_1}$ are the overall rates of the common and individual messages, respectively.

\noindent \textbf{Codebook Generation:} Fix a measure $P_{S,U,V,X_1,X_2,Y} \in \mc P$. Fix $\epsilon > 0$, $\eta_c > 0$, $\eta_1 > 0$, $\hat{\eta} > 0$, $\delta > 1$ and denote $M_c = 2^{nB[R_c-\eta_c\epsilon]}$, $M_1 = 2^{nB[R_1-\eta_1\epsilon]}$, $\hat{M} = 2^{n[\hat{R}+\hat{\eta}\epsilon]}$ and $J=2^{n[I(U;S|V,X_2)+\delta\epsilon]}$.

\noindent We randomly and independently generate a codebook for each block.

\begin{itemize}
\item[1)] For each block $i$, $i=1,\hdots,B$, we generate $M_c\hat{M}$ independent and identically distributed (i.i.d.) codewords $\dv x_{2,i}(w_c,t'_i)$ indexed by $w_c=1,\hdots,R_c$, $t'_i=1,\hdots,\hat{M}$, each with i.i.d. components drawn according to $P_{X_2}$.
\item[2)] For each block $i$, for each codeword $\dv x_{2,i}(w_c,t'_i)$,  we generate $\hat{M}$ i.i.d. codewords $\dv v_i(w_c,t'_i,t_i)$ indexed by $t_i=1,\hdots,\hat{M}$, each with i.i.d. components drawn according to $P_{V|X_2}$.
\item[3)]  For each block $i$, for each codeword $\dv x_{2,i}(w_c,t'_i)$, for each codeword $\dv v_i(w_c,t'_i,t_i)$, we generate a collection of $JM_1$ i.i.d. codewords $\{\dv u_i(w_c,t'_i,t_i,w_1,j_i)\}$ indexed by $w_1=1,\hdots,M_1$, $j_i=1,\hdots,J$, each with i.i.d. components draw according to $P_{U|V,X_2}$.
\end{itemize}

\textbf{Encoding:} Suppose that a common message $W_c=w_c$ and an individual message $W_1=w_1$ are to be transmitted. As we mentioned previously, $w_c$ and $w_1$ will be sent over \textit{all} blocks. We denote by $\dv s[i]$ the state affecting the channel in block $i$, $i=1,\hdots,B$. For convenience, we let $\dv s[0]=\emptyset$ and $t_{-1}=t_0=1$ (a default value). The encoding at the beginning of block $i$, $i=1,\hdots,B$, is as follows.

\noindent Encoder $2$, which has learned the state sequence $\dv s[i-1]$, knows $t_{i-2}$ and looks for a compression index $t_{i-1} \in [1:\hat{M}]$ such that $\dv v_{i-1}(w_c,t_{i-2},t_{i-1})$ is strongly jointly typical with $\dv s[i-1]$ and $\dv x_{2,i-1}(w_c,t_{i-2})$. If there is no such index or the observed state $\dv s[i-1]$ is not typical, $t_{i-1}$ is set to $1$ and an error is declared. If there is more than one such index $t_{i-1}$, choose the smallest. Encoder 2 then transmits the vector $\dv x_{2,i}(w_c,t_{i-1})$.

\noindent Encoder 1 obtains $\dv x_{2,i}(w_c,t_{i-1})$ similarly. It then finds the smallest compression index $t_i \in [1:\hat{M}]$ such that $\dv v_i(w_c,t_{i-i},t_i)$ is strongly jointly typical with $\dv s[i]$ and $\dv x_{2,i}(w_c,t_{i-1})$. Again, if there is no such index or the observed state $\dv s[i]$ is not typical, $t_i$ is set to $1$ and an error is declared. Next, Encoder 1 looks for the smallest $j_{i}$ such that $\dv u_i(w_c,t_{i-1},t_i,w_1,j_{i})$ is jointly typical with $\dv s[i]$  given $(\dv x_{2,i}(w_c,t_{i-1}),\dv v_i(w_c,t_{i-1},t_i))$. Denote this $j_{i}$ by $j^{\star}_{i}=j(\dv s[i],w_c,t_{i-1},t_i,w_1)$. If such $j^{\star}_{i}$ is not found, an error is declared and $j(\dv s[i],w_c,t_{i-1},t_i,w_1)$ is set to $j_{i}=J$. Encoder 1 then transmits a vector $\dv x_1[i]$ which is drawn i.i.d. conditionally given $\dv u_i(w_c,t_{i-1},t_i,w_1,j^{\star}_{i})$, $\dv s[i]$, $\dv v_i(w_c,t_{i-1},t_i)$ and $\dv x_{2,i}(w_c,t_{i-1})$ (using the conditional measure $P_{X_1|U,S,V,X_2}$ induced by \eqref{MeasureForCapacityRegionDiscreteMemorylessChannel}).

\textbf{Decoding:} At the end of the transmission, the decoder has collected all the blocks of channel outputs $\dv y[1],\hdots,\dv y[B]$.

\noindent \underline{\textit{Step (a):}} The decoder estimates message $w_c$ using \text{all} blocks $i=1,\hdots,B$, i.e., simultaneous decoding. It declares that $\hat{w}_c$ is sent if there exist $t^B=(t_1,\hdots,t_B) \in [1:\hat{M}]^{B}$, $w_1 \in [1:M_1]$ and $j^B=(j_{1},\hdots,j_{B}) \in [1:J]^B$ such that $\dv x_{2,i}(\hat{w}_c,t_{i-1})$, $\dv u_i(\hat{w}_c,t_{i-1},t_i,w_1,j_{i})$, $\dv v_i(\hat{w}_c,t_{i-1},t_i)$ and $\dv y[i]$ are jointly typical for all $i=1,\hdots,B$. One can show that the decoder obtains the correct $w_c$ as long as $n$ and $B$ are large and
\begin{align}
R_c + R_1 &\leq I(U,V,X_2;Y)-I(U,V,X_2;S).
\label{Constraint__On__SumRate}
\end{align}

\noindent \underline{\textit{Step (b):}} Next, the decoder  estimates message $w_1$ using again \text{all} blocks $i=1,\hdots,B$, i.e., simultaneous decoding. It declares that $\hat{w}_1$ is sent if there exist $t^B=(t_1,\hdots,t_B) \in [1:\hat{M}]^{B}$, $j^B=(j_{1},\hdots,j_{B}) \in [1:J]^B$ such that $\dv x_{2,i}(\hat{w}_c,t_{i-1})$, $\dv u_i(\hat{w}_c,t_{i-1},t_i,\hat{w}_1,j_{i})$, $\dv v_i(\hat{w}_c,t_{i-1},t_i)$ and $\dv y[i]$ are jointly typical for all $i=1,\hdots,B$. One can show that the decoder obtains the correct $w_1$ as long as $n$ and $B$ are large and
\begin{subequations}
\begin{align}
\label{Constraint__On__IndividualRate}
R_1 &\leq I(U;Y|V,X_2)-I(U;S|V,X_2)\\
R_1 &\leq I(U,V,X_2;Y)-I(U,V,X_2;S).
\end{align}
\end{subequations}
\hspace{7cm} \qed

\vspace{-0.3cm}

\subsection{Wyner-Ziv Binning With Nonunique Decoding is Optimal}\label{secIII_subsecA}

In the coding scheme of Theorem~\ref{Theorem__CapacityRegionDiscreteMemorylessChannel}, the state compression is standard, i.e., uses no Wyner-Ziv binning, the same message is sent in every block, and the decoding of the sent message is performed jointly using all blocks. Although of no benefit in the case of one relay, the combination of these three features was shown to be essential in achieving rates that are strictly larger than those offered by schemes based on Cover and El Gamal classic compress-and-forward scheme \cite{CG79} for certain networks with multiple relays in \cite{H-LKGC11}. That is, the coding scheme of \cite{H-LKGC11} outperforms Cover and El Gamal classic compress-and-forward for some multi-relay networks in \cite{H-LKGC11}. One can wonder whether the same holds for our model, i.e., whether schemes based on Cover and El Gamal classic compress-and-forward, i.e., block Markov encoding combined with Wyner-Ziv binning, fall short of achieving optimality for our model. In this paper, we show that the capacity region $\mc C$ as given by \eqref{CapacityRegionDiscreteMemorylessChannel} can be achieved alternatively with a coding scheme that we obtain by building upon and modifying Cover and El Gamal original compress-and-forward scheme. The modification consists essentially in 1) decoding block-by-block backwardly instead of block-by-block forwardly and 2) non-unique decoding of the compression indices. (In fact, by investigating more closely the converse proof of Theorem~\ref{Theorem__CapacityRegionDiscreteMemorylessChannel}, we will show later that 2) can be relaxed essentially without altering the capacity region). The following theorem states the result.

\vspace{0.3cm}

\begin{theorem}\label{Theorem__WynerZivBinningOptimality}
For the state-dependent multiaccess channel model that we study, there exists an optimal coding scheme that uses Wyner-Ziv binning for the state compression. That is, the capacity region $\mc C$ given by \eqref{CapacityRegionDiscreteMemorylessChannel} can also be achieved using a coding scheme in which the state compression is performed using Wyner-Ziv binning.
\end{theorem}

\vspace{0.3cm}

\textbf{Proof:} The achievability proof of Theorem~\ref{Theorem__WynerZivBinningOptimality} is based on a block-Markov coding scheme that combines carefully Gel'fand-Pinsker binning and Wyner-Ziv binning, and utilizes backward decoding with non-unique decoding of the compression indices. 

The transmission takes place in $B$ blocks. The common message $W_c$ is divided into $B$ blocks $w_{c,1},\hdots,w_{c,B}$ of $nR_c$ bits each, and the individual messages $W_1$ are divided into $B$ blocks $w_{1,1},\hdots,w_{1,B}$ of $nR_1$ bits each. For convenience, we let $w_{c,B}=w_{1,B}=1$ (a default value). We thus have $B_{W_c}=n(B-1){R_c}$, $B_{W_1}=n(B-1){R_1}$, $N=nB$, $R_{W_c}=B_{W_c}/N=R_c{\cdot}(B-1)/B$ and $R_{W_1}=B_{W_1}/N=R_1{\cdot}(B-1)/B$, where $B_{W_c}$ is the number of common message bits, $B_{W_1}$ is the number of individual message bits, $N$ is the number of channel uses and $R_{W_c}$ and $R_{W_1}$ are the overall rates of the common and individual messages, respectively. For fixed $n$, the average rate pair $(R_{W_c}, R_{W_1})$ over $B$ blocks can be made as close to $(R_c,R_1)$ as desired by making $B$ large.

\noindent \textbf{Codebook Generation:} Fix a measure $P_{S,U,V,X_1,X_2,Y} \in \mc P$. Fix $\epsilon > 0$ and denote $M_c = 2^{n[R_c-\eta_c\epsilon]}$, $M_1 = 2^{n[R_1-\eta_1\epsilon]}$, $M_0 = 2^{n[R_0+\eta_0\epsilon]}$, $\hat{M} = 2^{n[\hat{R}+\hat{\eta}\epsilon]}$, $J=2^{n[I(U;S|V,X_2)+\delta_U\epsilon]}$.

\begin{itemize}
\item[1)] We generate $M_cM_0$ independent and identically distributed (i.i.d.) codewords $\dv x_2(w_c,s)$ indexed by $w_c=1,\hdots,R_c$, $s=1,\hdots,M_0$, each with i.i.d. components drawn according to $P_{X_2}$.
\item[2)] For each codeword $\dv x_2(w_c,s)$,  we generate $\hat{M}$ independent and identically distributed (i.i.d.) codewords $\dv v(w_c,s,z)$ indexed by $z=1,\hdots,\hat{M}$, each with i.i.d. components drawn according to $P_{V|X_2}$.
\item[3)] For each codeword $\dv x_2(w_c,s)$, for each codeword $\dv v(w_c,s,z)$, we generate a collection of $JM_1$ i.i.d. codewords $\{\dv u(w_c,s,z,w_1,j)\}$ indexed by $w_1=1,\hdots,M_1$, $j=1,\hdots,J$, each with i.i.d. components draw according to $P_{U|V,X_2}$.
\item[4)] Randomly partition the set $\{1,\hdots,\hat{M}\}$ into $M_0$ cells $\mc C_s$, $s \in [1,M_0]$.
\end{itemize}

\textbf{Encoding:} Suppose that a common message $W_c=w_c$ and an individual message $W_1=w_1$ are to be transmitted. As we mentioned previously, message $w_c$ is divided into $B$ blocks $w_{c,1},\hdots,w_{c,B}$ and message $w_1$ is divided into $B$ blocks $w_{1,1},\hdots,w_{1,B}$, with $(w_{c,i},w_{1,i})$ the pair messages sent in block $i$. We denote by $\dv s[i]$ the channel state in block $i$, $i=1,\hdots,B$. For convenience, we let $\dv s[0]=\phi$ and $z_0=1$ (a default value), and $s_0$ the index of the cell containing $z_0$, i.e., $z_0 \in \mc C_{s_0}$ . The encoding at the beginning of the block $i$, $i=1,\hdots,B$, is as follows.

\noindent Encoder $2$, which has learned the state sequence $\dv s[i-1]$, knows $s_{i-2}$ and looks for a compression index $z_{i-1} \in [1,\hat{M}]$ such that $\dv v(w_{c,i-1},s_{i-2},z_{i-1})$ is strongly jointly typical with $\dv s[i-1]$ and $\dv x_2(w_{c,i-1},s_{i-2})$. If there is no such index or the observed state $\dv s[i-1]$ is not typical, $z_{i-1}$ is set to $1$ and an error is declared. If there is more than one such index $z_{i-1}$, choose the smallest. One can show that the probability of error of this event is arbitrarily small provided that $n$ is large and
\begin{align}
\hat{R} &> I(V;S|X_2).
\end{align}

Encoder 2 then transmits the vector $\dv x_2(w_{c,i},s_{i-1})$, where $s_{i-1}$ is such that $z_{i-1} \in \mc C_{s_{i-1}}$.

\noindent Encoder 1 obtains $\dv x_2(w_{c,i},s_{i-1})$ similarly. It then finds the smallest compression index $z_i \in [1,\hat{M}]$ such that $\dv v(w_{c,i},s_{i-1},z_i)$ is strongly jointly typical with $\dv s[i]$ and $\dv x_2(w_{c,i},s_{i-1})$. Again, if there is no such index or the observed state $\dv s[i]$ is not typical, $z_i$ is set to $1$ and an error is declared. Let $s_i \in [1,M_0]$ such that $z_i \in \mc C_{s_i}$. Next, Encoder 1 looks for the smallest $j_{i}$ such that $\dv u(w_{c,i},s_{i-1},z_i,w_{1,i},j_{i})$ is jointly typical with $\dv s[i]$, $\dv x_2(w_{c,i},s_{i-1})$ and $\dv v(w_{c,i},s_{i-1},z_i)$. Denote this $j_{i}$ by $j^{\star}_{i}=j(\dv s[i],w_{c,i},s_{i-1},z_i,w_{1,i})$. If such $j^{\star}_{i}$ is not found, an error is declared and $j(\dv s[i],w_{c,i},s_{i-1},z_i,w_{1,i})$ is set to $j_{i}=J$. Encoder 1 then transmits a vector $\dv x_1[i]$ which is drawn i.i.d. conditionally given $\dv s[i]$,  $\dv u(w_{c,i},s_{i-1},z_i,w_{1,i},j^{\star}_{i})$, $\dv v(w_{c,i},s_{i-1},z_i)$ and $\dv x_2(w_{c,i},s_{i-1})$ (using the conditional measure $P_{X_1|S,U,V,X_2}$ induced by  $P_{S,U,V,X_1,X_2,Y} \in \mc P$).

\textit{1) Decoding in Block $B-1$:}

The decoding of the pair $(w_{c,B-1},w_{1,B-1})$ is performed in four steps, as follows.

\noindent \underline{\textit{Step (a):}} The decoder knows $w_{c,B}=1$ and looks for the unique cell index $\hat{s}_{B-1}$ such that the vector $\dv x_2(w_{c,B},\hat{s}_{B-1})$ is jointly typical with $\dv y[B]$.  The decoding operation in this step incurs small probability of error as long as $n$ is sufficiently large and
\begin{align}
R_0 &< I(X_2;Y).
\label{DecodingOfCellIndex}
\end{align}

\noindent \underline{\textit{Step (b):}} The decoder now knows $\hat{s}_{B-1}$ (i.e., the index of the cell in which the compression index $z_{B-1}$ lies). It then decodes message $w_{c,B-1}$ by looking for the unique $\hat{w}_{c,B-1}$ such that $\dv x_2(\hat{w}_{c,B-1},s_{B-2})$,  $\dv v(\hat{w}_{c,B-1},s_{B-2},z_{B-1})$, $\dv u(\hat{w}_{c,B-1},s_{B-2},z_{B-1},w_{1,B-1},j_{B-1})$ and $\dv y[B-1]$ are jointly typical for some $s_{B-2} \in [1,M_0]$, $w_{1,B-1} \in [1,M_1]$, $j_{B-1} \in [1,J]$ and $z_{B-1} \in \mc C_{\hat{s}_{B-1}}$. One can show that the decoder obtains the correct $w_{c,B-1}$ as long as $n$ and $B$ are large and
\begin{align}
R_0+(\hat{R}-R_0)+ R_c + R_1 &\leq I(U,V,X_2;Y)-I(U;S|V,X_2).
\label{Constraint__On__SumRate}
\end{align}

\noindent \underline{\textit{Step (c):}} The decoder knows $\hat{w}_{c,B-1}$ and can again obtain the correct $s_{B-2}$ if $n$ is large and \eqref{DecodingOfCellIndex} is true. This is accomplished by looking for the unique $\hat{s}_{B-2}$ such that the vector $\dv x_2(\hat{w}_{c,B-1},\hat{s}_{B-2})$ is jointly typical with $\dv y[B-1]$.

\noindent \underline{\textit{Step (d):}} Finally, the decoder, which now knows message  $\hat{w}_{c,B-1}$ and the cell index $\hat{s}_{B-2}$ (but not the exact compression index $z_{B-1}$), estimates $w_{1,B-1}$ using $\dv y[B-1]$. It declares that $\hat{w}_{1,B-1}$ was sent if there exists a unique $\hat{w}_{1,B-1}$ such that $\dv x_2(\hat{w}_{c,B-1},\hat{s}_{B-2})$, $\dv v(\hat{w}_{c,B-1},\hat{s}_{B-2},z'_{B-1})$, $\dv u(\hat{w}_{c,B-1},\hat{s}_{B-2},z'_{B-1},\hat{w}_{1,B-1},j_{B-1})$ and $\dv y[B-1]$ are jointly typical for some $z'_{B-1} \in \mc C_{\hat{s}_{B-1}}$ and $j_{B-1} \in [1,J]$.
\begin{itemize}
\item If $z'_{B-1} = z_{B-1}$, the decoder finds the correct $w_{1,b-1}$ for sufficiently large $n$ if
\begin{align}
R_1 &\leq I(U;Y|V,X_2)-I(U;S|V,X_2).
\label{Constraint1__On__IndividualRate}
\end{align}
\item If $z'_{B-1} \neq z_{B-1}$, the decoder finds the correct $w_{1,b-1}$ for sufficiently large $n$ if
\begin{align}
(\hat{R}-R_0)+R_1 &\leq I(U,V;Y|X_2)-I(U;S|V,X_2).
\label{Constraint2__On__IndividualRate}
\end{align}
\end{itemize}

\textit{2) Decoding in Block $b$, $b=B-1,B-2,\hdots,2$:}

Next, for $b$ ranging from $B-1$ to $2$, the decoding of the pair $(w_{c,b-1},w_{1,b-1})$ is performed similarly, in five steps, by using the information $\dv y[b]$ received in block $b$ and the information $\dv y[b-1]$ received in block $b-1$. More specifically, this is done as follows.

\noindent \underline{\textit{Step (a):}} The decoder knows $w_{c,b}$ and looks for the unique cell index $\hat{s}_{b-1}$ such that the vector $\dv x_2(w_{c,b},\hat{s}_{b-1})$ is jointly typical with $\dv y[b]$. The decoding error in this step is small for sufficiently large $n$ if \eqref{DecodingOfCellIndex} is true.

\noindent \underline{\textit{Step (b):}} The decoder knows $\hat{s}_{b-1}$ and decodes message $w_{c,b-1}$ from $\dv y[b]$. It looks for the unique $\hat{w}_{c,b-1}$ such that $\dv x_2(\hat{w}_{c,b-1},s_{b-2})$,  $\dv v(\hat{w}_{c,b-1},s_{b-2},z_{b-1})$, $\dv u(\hat{w}_{c,b-1},s_{b-2},z_{b-1},w_{1,b-1},j_{b-1})$ and $\dv y[b-1]$ are jointly typical for some $s_{b-2} \in [1,M_0]$, $w_{1,b-1} \in [1,M_1]$, $j_{b-1} \in [1,J]$ and $z_{b-1} \in \mc C_{\hat{s}_{b-1}}$. One can show that the decoding error in this step is small for sufficiently large $n$ if \eqref{Constraint__On__SumRate} is true.

\noindent \underline{\textit{Step (c):}} The decoder knows $\hat{w}_{c,b-1}$ and obtains $\hat{s}_{b-2}$ by looking for the unique $\hat{s}_{b-2}$ such that the vector $\dv x_2(\hat{w}_{c,b-1},\hat{s}_{b-2})$ is jointly typical with $\dv y[b-1]$. For sufficiently large $n$, the decoder obtains the correct $s_{b-2}$ with high probability if \eqref{DecodingOfCellIndex} is true.

\noindent \underline{\textit{Step (d):}} Finally, the decoder, which now knows message  $\hat{w}_{c,b-1}$ and the cell index $\hat{s}_{b-2}$ (but not the exact compression index $z_{b-1}$), estimates message $w_{1,b-1}$ using $\dv y[b-1]$. It declares that $\hat{w}_{1,b-1}$ was sent if there exists a unique $\hat{w}_{1,b-1}$ such that $\dv x_2(\hat{w}_{c,b-1},\hat{s}_{b-2})$, $\dv v(\hat{w}_{c,b-1},\hat{s}_{b-2},z'_{b-1})$, $\dv u(\hat{w}_{c,b-1},\hat{s}_{b-2},z'_{b-1},\hat{w}_{1,b-1},j_{b-1})$ and $\dv y[b-1]$ are jointly typical for some $z'_{b-1} \in \mc C_{\hat{s}_{b-1}}$ and $j_{b-1} \in [1,J]$.
\begin{itemize}
\item If $z'_{b-1} = z_{b-1}$, the decoder finds the correct $w_{1,b-1}$ for sufficiently large $n$ if \eqref{Constraint1__On__IndividualRate} is true.
\item If $z'_{b-1} \neq z_{b-1}$, the decoder finds the correct $w_{1,b-1}$ for sufficiently large $n$ if \eqref{Constraint2__On__IndividualRate} is true.
\end{itemize}

Applying Fourier-Motzkin Elimination (FME) to project out  $R_0$ and $\hat{R}$ from \eqref{DecodingOfCellIndex},\eqref{Constraint__On__SumRate}, \eqref{Constraint1__On__IndividualRate} and \eqref{Constraint2__On__IndividualRate}, we get the desired result \eqref{CapacityRegionDiscreteMemorylessChannel}. \hspace{4cm} \qed

\vspace{-0.1cm}

\subsection{Non-unique Decoding is Not Needed}\label{secIII_subsecC}

As we mentioned previously, the coding scheme  of Theorem~\ref{Theorem__WynerZivBinningOptimality} shares elements with Cover and El Gamal original compress-and-forward \cite[Theorem 7]{CG79}; but differs from it mainly in two aspects. First, it uses backward decoding instead of the forward decoding of \cite{CG79}; and, second,  unlike \cite{CG79} it does not require unique decoding of the compression indices. The second aspect is essential for getting the \textit{same} rate expression as in \eqref{CapacityRegionDiscreteMemorylessChannel}, with no additional constraints. However, as we will see shortly in the corollary that will follow, one can modify the coding scheme of Theorem~\ref{Theorem__WynerZivBinningOptimality} in a way to get the compression indices decoded uniquely and \textit{still} get the capacity region, at the expense of slightly larger $|\mc V|$ and larger  $|\mc U|$. The key element is the observation that the constraint introduced by getting the compression index decoded, i.e., 
\begin{align}
I(V;S|X_2)-I(V;Y|X_2) &\leq I(X_2;Y),
\label{ConstraintOuterBound__Form2}
\end{align}
or, equivalently,
\vspace{-0.2cm}
\begin{align}
I(V,X_2;Y)-I(V,X_2;S) &\geq 0,
\label{ConstraintOuterBound__Form1}
\end{align}
is also \textit{implicit} in the converse proof of Theorem~\ref{Theorem__CapacityRegionDiscreteMemorylessChannel}. That is, the auxiliary random variables  $U$ and $V$ of the converse proof of Theorem~\ref{Theorem__CapacityRegionDiscreteMemorylessChannel} satisfy \eqref{ConstraintOuterBound__Form1}.

\vspace{0.2cm}

\begin{theorem}\label{Corollary__EquivalentCharacterizationCapacityRegionDiscreteMemorylessChannel}
The coding scheme of Theorem~\ref{Theorem__WynerZivBinningOptimality} can be modified in a way to get the compression index decoded. The resulting coding scheme is optimal and achieves an equivalent characterization of the capacity region of the model that we study given by the set of all rate pairs $(R_c,R_1)$ such that
\vspace{-0.3cm}
\begin{align}
R_1 \: &\leq \: I(U;Y|V,X_2)-I(U;S|V,X_2) \nonumber\\
R_c+ R_1 \: &\leq \: I(U,V,X_2;Y)-I(U,V,X_2;S)
\label{EquivalentCharacterizationCapacityRegionDiscreteMemorylessChannel}
\end{align}
for some measure $(S,U,V,X_1,X_2,Y) \in \mc P$ and satisfying
\begin{align}
I(V,X_2;Y)-I(V,X_2;S) &\geq 0,
\label{ConstraintOuterBound}
\end{align}
where the auxiliary random variables $V$ and $U$ have their alphabets bounded as

\vspace{-0.3cm}

\begin{subequations}
\begin{align}
\label{BoundsOnCardinalityOfAuxiliaryRandonVariableV__EquivalentCharacterizationCapacityRegion__DiscreteMemorylessChannel}
&\hspace{1.5cm}|\mc V| \leq |\mc S||\mc X_1||\mc X_2|+2\\
&\hspace{1.5cm}|\mc U| \leq \Big(|\mc S||\mc X_1||\mc X_2|+2\Big)|\mc S||\mc X_1||\mc X_2|.
\label{BoundsOnCardinalityOfAuxiliaryRandonVariableU__EquivalentCharacterizationCapacityRegion__DiscreteMemorylessChannel}
\end{align}
\label{BoundsOnCardinalityOfAuxiliaryRandonVariables__EquivalentCharacterizationCapacityRegion__DiscreteMemorylessChannel}
\end{subequations}

\end{theorem}

\vspace{-0.2cm}

\textbf{Proof:} The coding scheme that we use for the proof of Theorem~\ref{Corollary__EquivalentCharacterizationCapacityRegionDiscreteMemorylessChannel} is very similar to that of Theorem~\ref{Theorem__WynerZivBinningOptimality}, but with unique decoding of the compression indices. (See \cite{ZPS12b}). 

\section*{Acknowledgement}

This work has been supported by the European Commission in the framework of the FP7 Network of Excellence in Wireless Communications (NEWCOM++). The work of S. Shamai was supported by the Philipson Fund for Electrical Power, via the Technion research authority.

\vspace{-0.1cm}

\bibliographystyle{IEEEtran}
\bibliography{paperISIT2012}

\begin{thebibliography}{10}
\providecommand{\url}[1]{#1}
\csname url@rmstyle\endcsname
\providecommand{\newblock}{\relax}
\providecommand{\bibinfo}[2]{#2}
\providecommand\BIBentrySTDinterwordspacing{\spaceskip=0pt\relax}
\providecommand\BIBentryALTinterwordstretchfactor{4}
\providecommand\BIBentryALTinterwordspacing{\spaceskip=\fontdimen2\font plus
\BIBentryALTinterwordstretchfactor\fontdimen3\font minus
  \fontdimen4\font\relax}
\providecommand\BIBforeignlanguage[2]{{%
\expandafter\ifx\csname l@#1\endcsname\relax
\typeout{** WARNING: IEEEtran.bst: No hyphenation pattern has been}%
\typeout{** loaded for the language `#1'. Using the pattern for}%
\typeout{** the default language instead.}%
\else
\language=\csname l@#1\endcsname
\fi
#2}}

\bibitem{LS10a}
A.~Lapidoth and Y.~Steinberg, ``The multiple access channel with causal and
  strictly causal side information at the encoders,'' in \emph{Proc. Int.
  Zurich Seminar on Communications (IZS)}, Zurich, Switzerland, Mar. 2010, pp.
  13--16.

\bibitem{D80}
G.~Dueck, ``Partial feedback for two-way and broadcast channels,'' \emph{Inf.
  Contr.}, vol.~46, pp. 1--15, 1980.

\bibitem{LS10b}
A.~Lapidoth and Y.~Steinberg, ``The multiple access channel with two
  independent states each known causally at one encoder,'' in \emph{Proc.
  {IEEE} Int. Symp. Information Theory}, Austin, TX, USA, Jun. 2010, pp.
  480--484.

\bibitem{LSY10}
M.~Li, O.~Simeone, and A.~Yener, ``Multiple access channels with states
  causally known at transmitters,'' \emph{Submitted for publication in IEEE
  Trans. Inf. Theory. Available at \url{http://arxiv.org/abs/1011.6639}}, 2010.

\bibitem{H-LKGC11}
S.~H. Lim, Y.-H. Kim, A.~E. Gamal, and S.-Y. Chung, ``Noisy network coding,''
  \emph{IEEE Trans. Inf. Theory}, vol.~57, pp. 3132--3152, May 2011.

\bibitem{WZ76}
A.~D. Wyner and J.~Ziv, ``The rate-distortion function for source coding with
  side information at the decoder,'' \emph{{IEEE} Trans. Inf. Theory}, vol.~22,
  pp. 1--10, Jan. 1976.

\bibitem{LS11a}
A.~Lapidoth and Y.~Steinberg, ``A note on multiple access channels with
  strictly causal state information,'' in \emph{available at
  \url{http://arxiv.org/abs/1106.0380}}, Jun. 2011.

\bibitem{ZPS11a}
A.~Zaidi, P.~Piantanida, and S.~{Shamai (Shitz)}, ``Multiple access channel
  with states known noncausally at one encoder and only strictly causally at
  the other encoder,'' in \emph{Proc. {IEEE} Int. Symp. Information Theory},
  Saint-Petersburg, Russia, 2011, pp. 2801--2805.

\bibitem{KH11a}
G.~Kramer and J.~Hou, ``On message lengths for noisy network coding,'' in
  \emph{Proc. {IEEE} Information Theory Workshop}, Paraty, Brasil, Oct. 2011.

\bibitem{WX11a}
X.~Wu and L.-L. Xie, ``On the optimal compressions in the compress-and-forward
  relay schemes,'' \emph{IEEE Trans. Inf. Theory, submitted for publication.
  Available at \url{http://arxiv.org/abs/1009.5959}}, Feb. 2011.

\bibitem{ZPS12b}
A.~Zaidi, P.~Piantanida, and S.~{Shamai (Shitz)}, ``Capacity region of multiple
  access channel with states known noncausally at one encoder and only strictly
  causally at the other encoder,'' \emph{IEEE Trans. Inf. Theory, submitted for
  publication, available at \url{http://arxiv.org/abs/1201.3278}}, 2012.

\bibitem{GP80}
S.~I. Gel'fand and M.~S. Pinsker, ``Coding for channel with random
  parameters,'' \emph{Problems of Control and Information Theory}, vol.~9, pp.
  19--31, 1980.

\bibitem{CG79}
T.~M. Cover and A.~{El Gamal}, ``Capacity theorems for the relay channel,''
  \emph{IEEE Trans. Inf. Theory}, vol. IT-25, pp. 572--584, Sep. 1979.

\end{thebibliography}
\end{document}